\documentclass[aps,prl, showpacs, preprint,superscriptaddress]{revtex4-1}

\usepackage{multirow}
\usepackage{graphicx}
\usepackage{amssymb,amsmath}
\usepackage{bm} 
\usepackage{float}
\usepackage{xcolor}
\definecolor{midnightblue}{cmyk}{1,1,0,0.1}
\definecolor{forestgreen}{cmyk}{0.75,0,1,0.5}

\usepackage{hyperref}
\hypersetup{
    bookmarks=true,                 
    unicode=false,                     
    pdftoolbar=true,                   
    pdfmenubar=true,                
    pdffitwindow=false,              
    pdfstartview={FitH},              
    pdftitle={My title},                  
    pdfauthor={Author},              
    pdfsubject={Subject},            
    pdfcreator={Creator},             
    pdfproducer={Producer},       
    pdfkeywords={keyword1} {key2} {key3}, 
    pdfnewwindow=true,             
    colorlinks=true,                      
    linkcolor=midnightblue,          
    citecolor=magenta,                
    filecolor=midnightblue,           
    urlcolor=midnightblue,            
}

\begin{document}

\title{Computational Study of the Magnetic Structure of Na$_2$IrO$_3$} 

\author{Kaige Hu}
\affiliation{International Center for Quantum Materials, School of Physics, Peking University, Beijing 100871, China}
\affiliation{Collaborative Innovation Center of Quantum Matter, Beijing 100871, China}

\author{Fa Wang}
\email{wangfa@pku.edu.cn}
\affiliation{International Center for Quantum Materials, School of Physics, Peking University, Beijing 100871, China}
\affiliation{Collaborative Innovation Center of Quantum Matter, Beijing 100871, China}

\author{Ji Feng}
\email{jfeng11@pku.edu.cn}
\affiliation{International Center for Quantum Materials, School of Physics, Peking University, Beijing 100871, China}
\affiliation{Collaborative Innovation Center of Quantum Matter, Beijing 100871, China}

\begin{abstract}
The magnetic structure of honeycomb iridate Na$_2$IrO$_3$ is of paramount importance to its exotic properties. The magnetic order is established experimentally to be zigzag antiferromagnetic. However, the previous assignment of ordered moment to the $\bm{a}$-axis is tentative. 
We examine the magnetic structure of Na$_{2}$IrO$_{3}$ using first-principles methods. Our calculations reveal that total energy is minimized when the zigzag antiferromagnetic order is magnetized along $\bm{g}\approx\bm{a}+\bm{c}$. Such a magnetic configuration is explained by adding anisotropic interactions to the nearest-neighbor Kitaev-Heisenberg model.  Spin-wave spectrum is also calculated, where the calculated spin gap of $10.4$ meV can in principle be measured by future inelastic neutron scattering experiments. Finally we emphasize that our proposal  is consistent with all known experimental evidence, including the most relevant resonant x-ray magnetic scattering measurements [X. Liu \emph{et al.} {Phys. Rev. B} \textbf{83}, 220403(R) (2011)]. 
\end{abstract}


\maketitle

The 5\emph{d} iridium-based transition metal oxides display very rich, interesting properties owing to the interplay between spin-orbit coupling, electron correlation, and crystal-field splitting \cite{Kim2008,Kim2009,Pesin2010,Jackeli2009,Chaloupka2010,Wan2011}.
In particular, A$_{2}$IrO$_{3}$ (A = Na, Li) have attracted special attentions \cite{Jackeli2009,Chaloupka2010, Singh2010, Kimchi2011, Liu2011, Choi2012, Ye2012,Lovesey2012, Singh2012, Comin2012, Chaloupka2013}, whose structure may be characterized as layered honeycomb lattices of Ir. The octahedrally coordinated Ir$^{\text{4+}}$ ion is suggested to possess an effective $j_{\text{eff}}=1/2$ pseudospin and  the edge-sharing oxygen octahedron structure is proposed to realize the Kitaev model \cite{Jackeli2009,Chaloupka2010}.
As an exactly solvable quantum spin-1/2 system, the Kitaev model embodies Majorana fermion excitations and quantum spin liquid that have potential implication to quantum computing \cite{Kitaev2006}.
 Although experiments have shown that the magnetic structure of Na$_2$IrO$_3$ is not a spin liquid but zigzag antiferromagnetic (AFM) \cite{Choi2012,Ye2012}, the understanding of such a exotic magnetic structure will provide important clues to realizing Kitaev spin liquid in this
 family of materials.

It is crucial to point out that although zigzag AFM order is well established experimentally, the assignment of direction of the AFM moments, on the other hand, is not without ambiguity. The zigzag AFM order was first proposed by combining resonant x-ray magnetic scattering measurements and first-principles calculations, with the ordered moment assigned to the crystallographic $\bm{a}$-axis \cite{Liu2011}. In later experiments that confirmed the zigzag configuration with neutron scattering, the moment direction was inherited without further scrutiny \cite{Choi2012,Ye2012}. Apparently, two standing issues remain with  the magnetic structure of Na$_{2}$IrO$_{3}$.  First, the determination of magnetic moment direction is still far from conclusive.  There is inconsistency between the tentative experimental assignment and first principles calculations: previous calculations predicted that the zigzag configuration have lower total energy for magnetic moments along the $\bm{b}$-axis compared a configuration magnetized along the $\bm{a}$-axis \cite{Liu2011}. 
As the proposed Kitaev-like models hinges upon anisotropic interactions, the determination of actual direction of the AFM order parameter is clearly critical for establishing reliable microscopic understanding of the low energy excitations in this compound. Second, and indeed, the microscopic models of Na$_{2}$IrO$_{3}$ are subject to controversy. For the Kitaev-Heisenberg (KH) model, which has various modifications and has been mostly adopted in literature \cite{Chaloupka2010,Chaloupka2013,Kimchi2011,Singh2012,Choi2012,Kim2012}, the isotropic Heisenberg interactions do not lead to a special preferred moment direction while the anisotropic Kitaev interactions make the moments prefer the cubic $\hat{\bm{z}}$-axis of the local IrO$_6$ octahedron (to be discussed later).  
Several recent studies \cite{Katukuri2014,Rau2014prl,Yamaji2014,Rau2014arxiv,Sizyuk2014arxiv, Chaloupka15arxiv} analysed the necessity to adding anisotropic interactions to the KH Hamitonian, which was expected to stabilize the zigzag configuration.
However, the puzzle of magnetic moment direction assignment remains. 

In this Letter, we employ the first principles method to examine the energetics of Na$_{2}$IrO$_{3}$, sampling a wide range of magnetic order with different moment alignments. Our calculations show that the ground state is attained in the zigzag AFM structure, with a moment direction  $\bm{g}\approx\bm{a}+\bm{c}$. 
We further show that the first principles energies can be well fitted with a modified nearest-neighbor Kitaev-Heisenberg (nnKH) Hamiltonian of spin-1/2 by adding anisotropic interactions, in which the Kitaev term dominates. 
Based on this model, we derive a few experimentally accessible quantities, 
such as  the spin wave spectrum.  Finally, we clarify that this assignment of moment direction is also consistent with resonant x-ray magnetic scattering measurements \cite{Liu2011}.

\begin{figure}
\includegraphics[width=0.618\columnwidth]{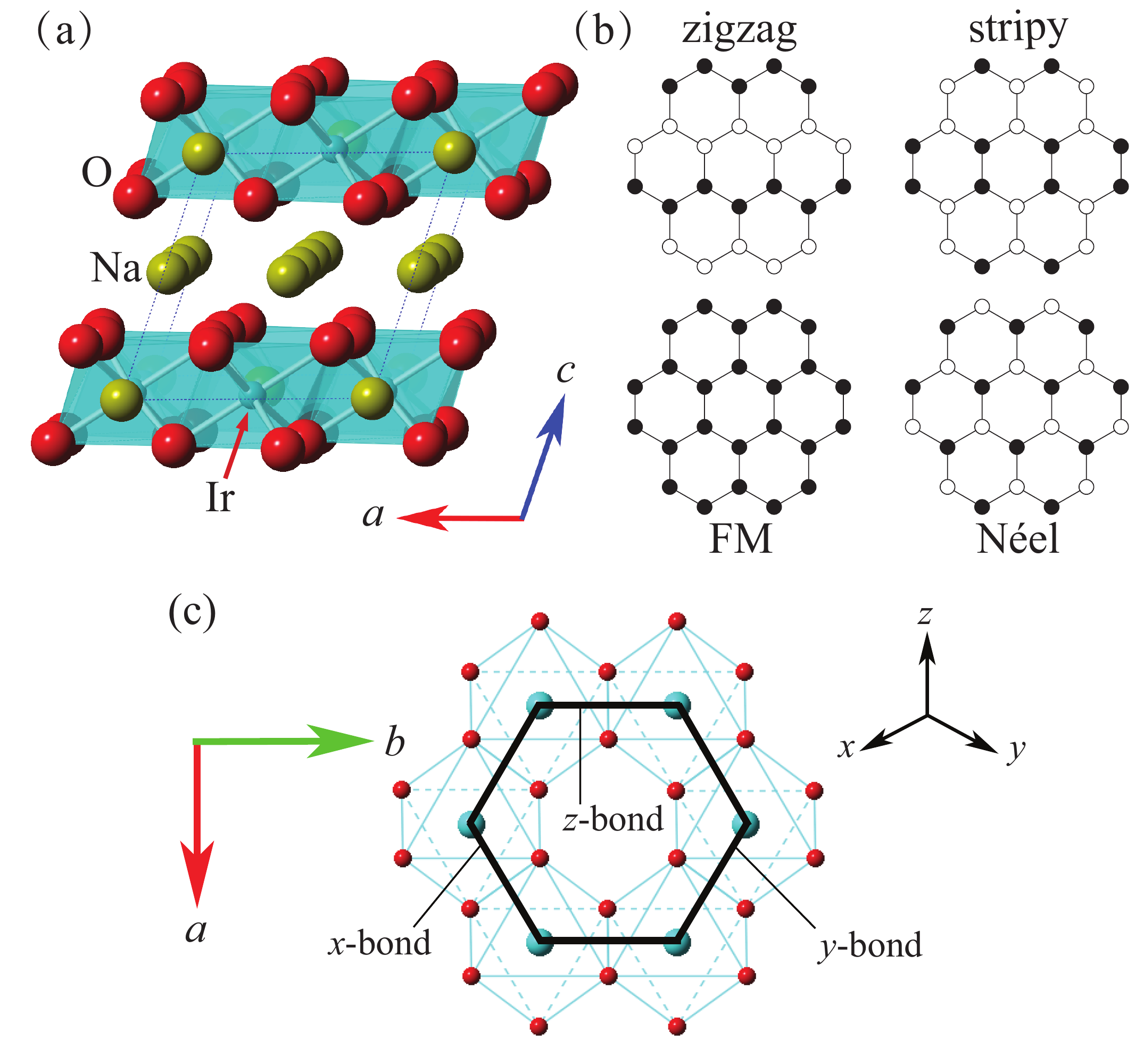}
\caption{(Color online) 
(a) The $C2/m$ crystal structure of Na$_{2}$IrO$_{3}$, viewed from slightly off the $\bm{b}$-direction. 
(b) Four different types of magnetic order. White and black circles denote up and down spins, respectively.  (c) Three different types of nearest-neighbor Ir-Ir bonds. The iridium honeycomb plane is perpendicular to the cubic direction [111].
\label{fig:1}}
\end{figure}

Na$_2$IrO$_3$ is a layered compound (space group $C2/m$), in which Ir ions are located at the center of edge-sharing octahedra formed by oxygen anions (Fig.~\ref{fig:1}(a)) \cite{Choi2012,Ye2012}.  Thus, Ir ions form a honeycomb lattice within each layer. 
Each Ir$^{4+}$ ion has five 5\emph{d} electrons, occupying $t_{2g}$ orbitals of the ideal octahedral crystal field assuming the oxygen octahedra remain regular. Owing to the strong spin-orbits coupling (SOC), the six $t_{2g}$ spin-orbitals are further separated into two manifolds with, respectively, $j_{\text{eff}}=3/2$ and  $j_{\text{eff}}=1/2$ \cite{Jackeli2009}.
The bands mainly composed of the $j_{\text{eff}}=3/2$ states are fully filled, while the spin-orbit-coupled $j_{\text{eff}}=1/2$ states are half filled, a keen observation that lead Khalliulin \emph{et al} \cite{Chaloupka2010} to relate this material to the Kitaev's spin-1/2 model, with an additional Heisenberg-type interactions, in what is called the Kitaev-Heisenberg models. Four types of magnetic order of the $j_{\text{eff}}=1/2$ pseudospin are shown in Fig.~\ref{fig:1}(b), namely, ferromagnetic (FM), N\'{e}el AFM, stripy AFM, and zigzag AFM. 
Four theoretical models are proposed to account for the magnetic order in Na$_2$IrO$_3$:
(1) the nnKH model which only includes nearest-neighbor interaction between Ir atoms \cite{Chaloupka2010,Chaloupka2013}, 
(2) the KH-$J_{2}$-$J_{3}$ model which also includes the second and third nearest neighbor Heisenberg hopping $J_2$ and $J_3$ between Ir atoms \cite{Kimchi2011,Choi2012,Kim2012,Singh2012},
(3) the \emph{modified} nnKH model which includes additional anisotropic interactions besides Kitaev terms and Heisenberg terms \cite{Katukuri2014,Rau2014prl,Yamaji2014,Rau2014arxiv,Sizyuk2014arxiv}, and
(4) the quasimolecular orbital model \cite{Mazin2012,Mazin2013,Foyevtsova2013}. 
The KH models are based upon a local moment picture while the quasimoleular orbital model an itinerant picture. 

\begin{figure}
\includegraphics[width=0.75\columnwidth]{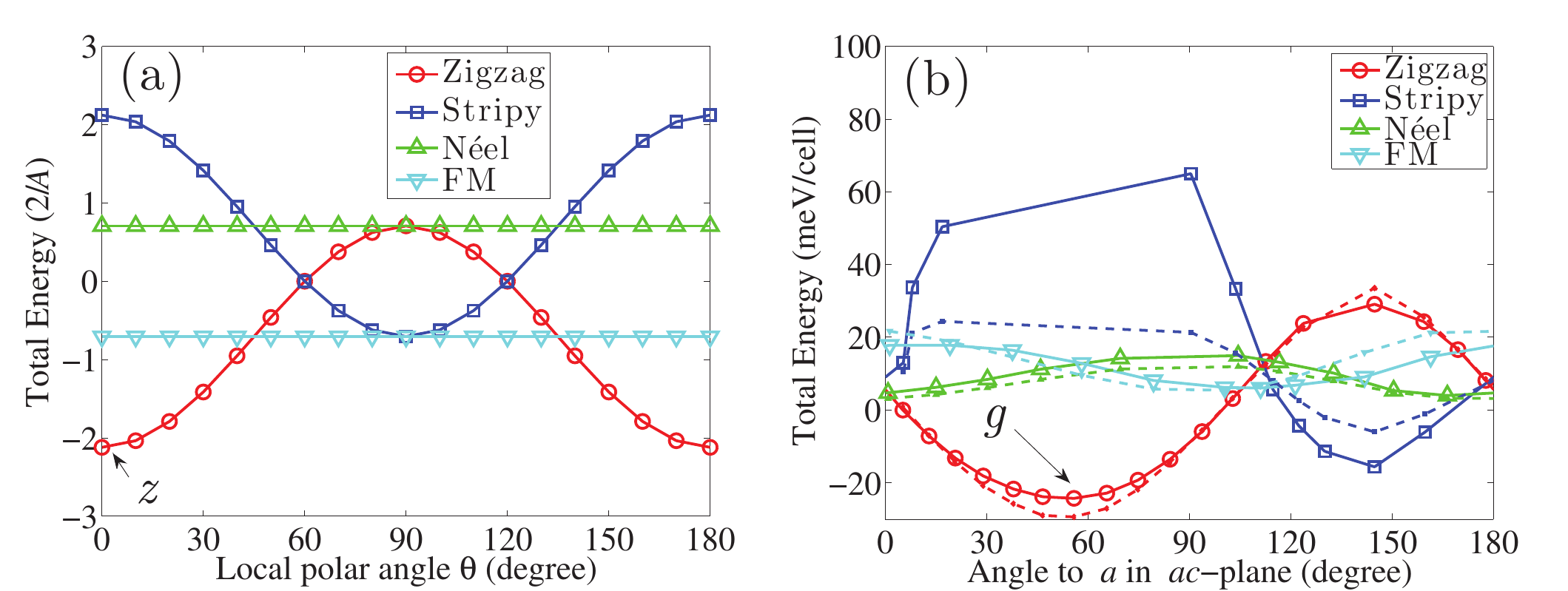}
\caption{(Color online) 
(a) Anisotropic energy of the KH model in the $\zeta=3\pi/4$ zigzag state. 
The angle $\theta$ is the polar angle in the local spherical coordinates of the IrO$_6$ octahedron. 
(b) Anisotropic energy in the $ac$-plane by first-principles calculations (solid lines) versus the total moment for the experimental structure of Na$_{2}$IrO$_{3}$ . Angles are measured from the $\bm{a}$ axis. 
The energy of the zigzag order with the moment along the $\bm{a}$ axis is set to be 0. Corresponding fitted curves are also shown (dash lines).
\label{fig:2}}
\end{figure}

The nnKH model is the simplest model that produce the zigzag AFM ground state, with the following Hamiltonian \cite{Chaloupka2013}:
\begin{equation}
H=\sum_{\gamma=x,y,z}\sum_{<i j>\in \gamma}\left(2KS_{i}^{\gamma}S_{j}^{\gamma}+J\mathbf{S}_{i}\cdot\mathbf{S}_{j}\right),\label{eq:nnKH}
\end{equation}
where the first term is the strongly anisotropic Kitaev interaction \cite{Kitaev2006} ($\gamma=x,y,z$ refers to the three nn bonds and also the three local axes along the Ir-O bonds of the IrO$_6$ octahedron shown in Fig.~\ref{fig:1}c), and the second one is the Heisenberg term. Eq.~(\ref{eq:nnKH}) can be rewritten as \cite{Chaloupka2013}  
 $H=\sum_{\gamma}\sum_{<i j>}A\left(2\sin\zeta S_{i}^{\gamma}S_{j}^{\gamma}+\cos\zeta\mathbf{S}_{i}\cdot\mathbf{S}_{j}\right)$, where $A=\sqrt{K^{2}+J^{2}}$ is a positive energy scale and the variety of the "phase" angle $\zeta$ tune the sign and relative strength of the Kitaev type and the Heisenberg type contributions in the parameter space. 
The anistropic energy for the four possible magnetic patterns, i.e., FM, N\'{e}el, stripy and zigzag, can be expressed as $E_{\textrm{zigzag}}=\frac{A}{2}\left(\cos\zeta-2\sin\zeta\cos\left(2\theta\right)\right)$, $E_{\textrm{stripy}}=-E_{\textrm{zigzag}}$, $E_{\textrm{FM}}=\frac{A}{2}\left(3\cos\zeta+2\sin\zeta\right)$, and $E_{\textrm{N\'{e}el}}=-E_{FM}$, where $\theta$ is the polar angle in the local spherical coordinates of the IrO$_{6}$ octahedron. 
Fig.~\ref{fig:2}(a) shows that when the zigzag magnetic order is the ground state ($\zeta=3\pi/4$ in the figure), the magnetic moment points along the local $\hat{\bm{z}}$-direction. This conclusion is consistent with the assumptions in Ref.~\cite{Choi2012}. The  KH-$J_{2}$-$J_{3}$ model should produce the same qualitative conclusion on the anisotropic energy since the Heisenberg terms are isotropic.
 
Motivated by the foregoing analysis, we perform detailed investigations on the anisotropic energy by non-collinear relativistic density functional theory, as implemented in Vienna \emph{ab-initio} simulation package \cite{Kresse1993,Kresse1996}. 
The experimental structure of Na$_2$IrO$_3$ is adopted \cite{Choi2012}.
The magnetic unit cell is chosen the same as the crystal unit cell, containing one layer of four Ir atoms, which is consistent with the consideration in the KH model.
The projector-augmented wave potentials \cite{Kresse1999} with a plane-wave cutoff of 500 eV is employed. 
We use the Monkhorst-Pack \emph{k}-point meshes \cite{Monkhorst1976} of $6\times4\times6$ per magnetic unit cell to perform the Brillouin zone summation.
We set $U=1.7$ eV, and $J=0.6$ eV \cite{Marel1988}, which corresponds to $U_{\text{eff}}=U-J=1.1$ eV \cite{Dudarev1998}.
Such choice of $U_{\text{eff}}$ result in a band gap of 341 meV for the ground zigzag state, consistent with the experimentally measured values (340 meV in Ref.~\cite{Comin2012}).
We perform complete self-consistent calculations with the spin-orbit coupling interaction. 
To survey the potential energy surface of magnetization, the spin magnetic moment is constrained in specified directions while the magnitude is optimized.

Figure \ref{fig:2}(b) shows the anisotropic total energies of the four magnetic configurations in the $ac$-plane for the experimental structure of Na$_{2}$IrO$_{3}$.  
The horizontal axis is the the angle between the total moment and the $\bm{a}$-axis, where the total moment is the summation of the spin moment and the orbital moment. 
Surprisingly, although the zigzag state is indeed the ground state, the total moment points along neither the cubic $\hat{\bm{z}}$-axis suggested by the KH model, nor the crystallographic $\bm{a}$-axis suggested in Ref.~\cite{Liu2011}. 
The energy in the $ac$-plane reaches its minimum value when the total moment points to the direction $\bm{g}\approx\bm{a}+\bm{c}$, which forms an angle of 55$^\circ$ with the $\bm{a}$-axis (see Fig.~\ref{fig:3}(a), where the AFM coupling between Ir honeycomb planes will be discussed in Supplemental Material \cite{Hu2015}). 
The $\bm{g}$-configuration's energy is significantly lower than the $\bm{a}$-configuration by about 24 meV per cell (4 Ir). 
To present the $\bm{g}$-direction more clearly, Fig.~\ref{fig:3}(b) shows the relative relations of the total moment direction $\bm{g}$, the crystallographic axes $\bm{a}$, $\bm{b}$ and $\bm{c}$, and the local axes of the IrO$_{6}$ octahedron $\hat{\bm{x}}$, $\hat{\bm{y}}$ and $\hat{\bm{z}}$ which connect an Ir atom to one of the nearest O atoms. 
It is interesting to note that $\bm{g}=2a_{0}\left(1,1,0\right)$, where $a_0$ is the Ir-O bond length, {i.e.}, $\bm{g}$ is a high-symmetry direction of the local IrO$_6$ octahedron, [110]. 
The $\bm{g}$-direction is located in the cubic $xy$-plane and points to the middle of one O-O edge. 
The anisotropic energy reaches its maximum value in the $ac$-plane when the total moment points to the cubic $\hat{\bm{z}}$-axis. 

\begin{figure}
\includegraphics[width=0.618\columnwidth]{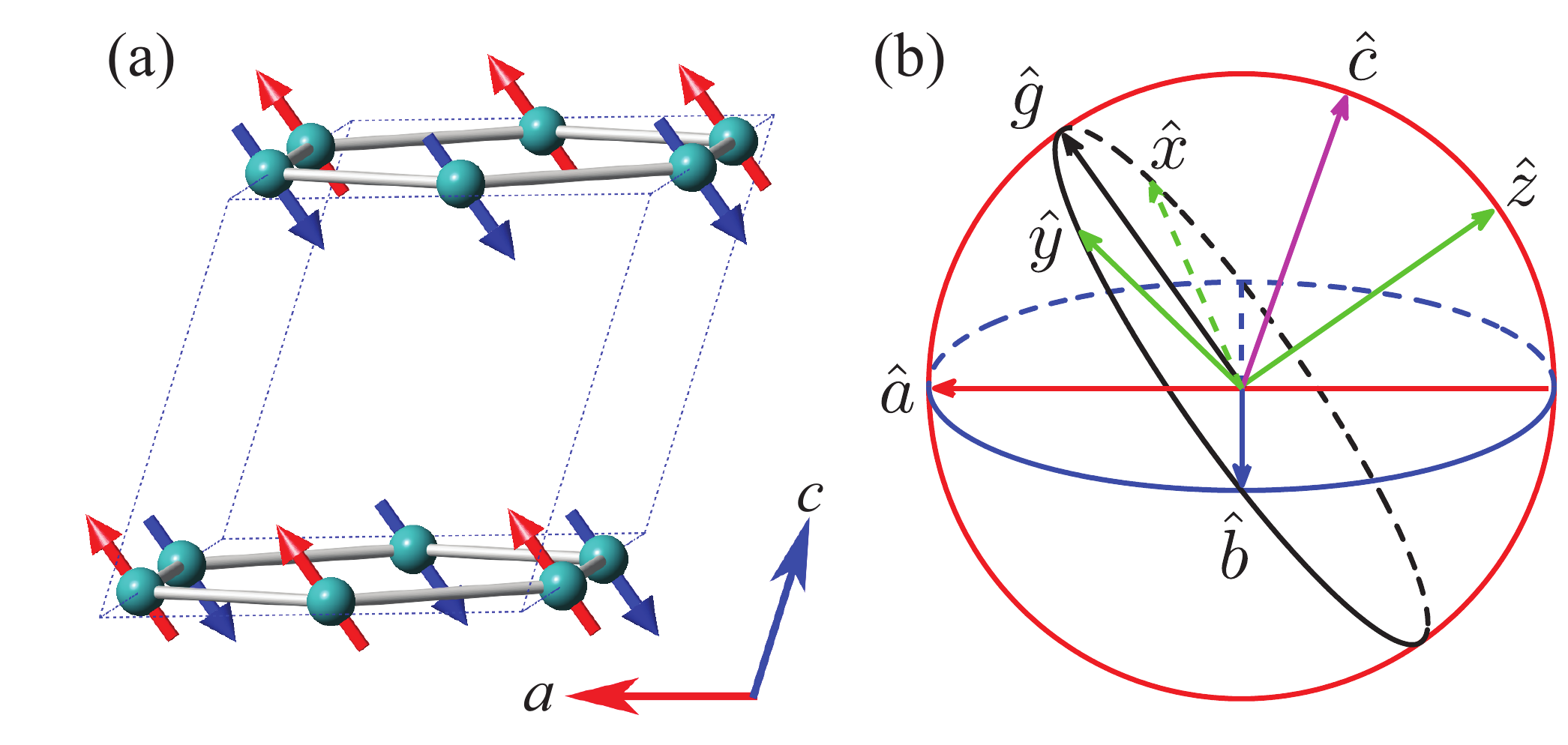}
\caption{(Color online) 
(a) The Ir honeycomb structure of Na$_{2}$IrO$_{3}$ and the zigzag magnetic order of the ground magnetic state. 
(b) Relative relations of the local IrO$_{6}$ axes $\hat{\bm{x}}$, $\hat{\bm{y}}$ and $\hat{\bm{z}}$, the crystallographic axes $\bm{a}$, $\bm{b}$, and $\bm{c}$, and also the moment direction $\bm{g}$.
\label{fig:3}}
\end{figure}

\begin{figure}
\includegraphics[width=.75\columnwidth]{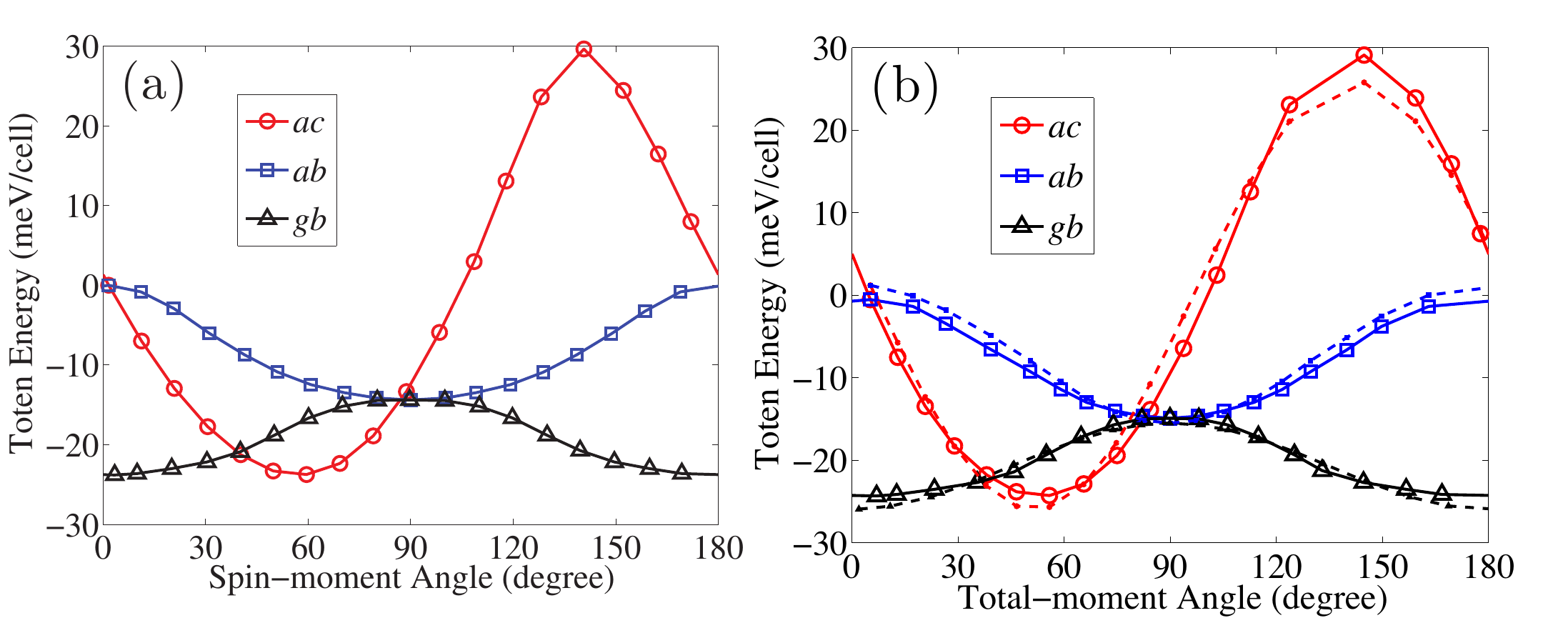}
\caption{(Color online) 
(a) Anisotropic energy by first-principles calculations where the angle of the spin moment is scanned in three different planes: $ac$, $ab$, and $gb$. 
Angles are measured from the $\bm{a}$-axis to the moment direction for the $ac$- and $ab$-plane, and from the $\bm{g}$-direction for the $gb$-plane, respectively. 
(b) Correspoinding anisotrpic energy by first-principles calculations versus the total moment direction (solid lines). Corresponding fitted curves by the modified nnKH model are also shown (dash lines).
\label{fig:4}}
\end{figure}

Figure \ref{fig:4} further confirms that the $\bm{g}$-direction is  actually the moment direction of the ground zigzag state,  consistent with resonant x-ray magnetic scattering measurement suggesting that  magnetic moments lie  in the $ac$-plane \cite{Liu2011,Lovesey2012}. On the other hand, the $\hat{\bm{z}}$-direction corresponds to the global maximum energy. To show this, the anisotropic energy is computed with the spin moment touring in three different planes: the $ac$, $ab$, and $gb$-plane (see Fig.~\ref{fig:3}(b)). The scanned moment angles are measured from the $\bm{a}$-direction for the $ac$- and $ab$-plane, and from the $\bm{g}$-direction for the $gb$-plane, respectively. The horizontal axes are the angle of the spin moment in Fig.~\ref{fig:4}(a) and the total moment in Fig.~\ref{fig:4}(b), respectively. The spin and orbital moments are nearly collinear, with a mutual angle less than 15$^{\circ}$. As a consequence, the curves in Fig.~\ref{fig:4}(a) are similar to that in Fig.~\ref{fig:4}(b). For the zigzag configuration, the angle of the $\bm{g}$-direction relative to the $\bm{a}$-axis is about 60$^\circ$ for the spin moment and 55$^\circ$ for the total moment. The total moment is ideally located in the cubic $xy$-plane of the IrO$_{6}$ octahedron. When the moment points along the $\bm{a}$-axis, the energy is higher than that of both the $\bm{b}$- and $\bm{g}$-directions. The energy with the moment pointing along the $\bm{b}$-axis is a saddle point on the potential energy surface: it is the minimum in the $ab$-plane and the maximum in the $gb$-plane (i.e, the cubic $xy$-plane). 
It is higher than the ground energy (that of the $\bm{g}$ direction) by 9.4 meV per cell. Therefore we conclude that the $gb$-plane (the cubic $xy$-plane) is an ``easy" plane.

Now we turn to the model explanation of the moment assignment. 
The prediction of magnetic moments along the $\hat{\bm{z}}$-axis indicates that the KH model is clearly inadequate.
Here we show that the $\bm{g}$-direction assignment of magnetic moment can be explained a modified nnKH model with additional anisotropic interactions, where the parameters can be fitted from the first-principles energies. 
The generalized model is described as
\begin{equation}
H=\sum_{\alpha,\beta=x,y,z}\sum_{<i j>}S_i^\alpha J_{ij}^{\alpha\beta}S_j^\beta,\label{eq:modified_nnKH}
\end{equation}
where the $3\times 3$ matrices
$J_{ij}$ on $x$,$y$,$z$-bonds are
\begin{align*}
\begin{pmatrix}
J+2K & J_{\parallel\perp} & J_{\parallel\perp} \\
J_{\parallel\perp} & J & J_{\perp\perp} \\
J_{\parallel\perp} & J_{\perp\perp} & J
\end{pmatrix}
, & \quad
\begin{pmatrix}
J & J_{\parallel\perp} & J_{\perp\perp} \\
J_{\parallel\perp} & J+2K & J_{\parallel\perp} \\
J_{\perp\perp} & J_{\parallel\perp} & J
\end{pmatrix}
,\quad
\begin{pmatrix}
J & J_{\perp\perp} & J_{\parallel\perp} \\
J_{\perp\perp} & J & J_{\parallel\perp} \\
J_{\parallel\perp} & J_{\parallel\perp} & J+2K
\end{pmatrix}
, &
\end{align*}
respectively. 

The form of these anisotropic exchange interactions is fixed by the assumption of perfect honeycomb lattice symmetry ($D_{3d}$ symmetry at Ir sites), and has been reported before \cite{Rau2014arxiv}. The lower symmetry of real Na$_2$IrO$_3$ crystals will in principle produce more complex anisotropies \cite{Yamaji2014}, which we will however not consider in this work.
In fitting the energies we treat the (pseudo-)spins $S_i^a$ as classical vectors. 
This model can naturally explain the zigzag AFM ground state
without invoking further neighbor interactions.
It can also produce the local $[110]$ moment direction for zigzag state.
The fitted curves are plotted in Fig. \ref{fig:4}(b) (dash lines), with model parameters from the second column of Table I in Supplemental Material \cite{Hu2015}, where details of the fitting results are also presented. The fitting turns out to be quite good.

From the energy dependence of moment direction for the zigzag AFM state shown in Fig.~\ref{fig:4}(b), we can fit the Kitaev term coefficient $K$, and anisotropy terms $J_{\parallel\perp}$ and $J_{\perp\perp}$. The energies of other magnetic orders shown in Fig.~\ref{fig:2}(b) are required to fit the Heisenberg couplings. 
Note that although the modified nnKH model can explain the $\bm{g}$-direction moment assignment of the zigzag state,
more interactions are necessary to satisfy the condition for zigzag ground state.  
The fitted curves are plotted in Fig. \ref{fig:2}(b) (dash lines), with model parameters from the second column of Table III in Supplemental Material \cite{Hu2015}, where details of the fitting results are also presented. Our main conclusion is that the dominant interaction is ferromagnetic Kitaev term.

\begin{figure}
\includegraphics[width=0.618\columnwidth]{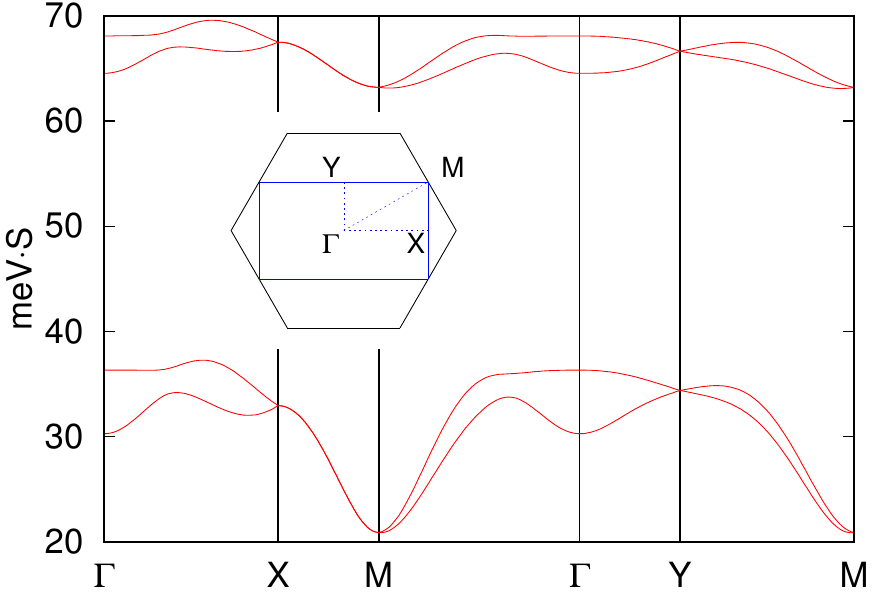}
\caption{(Color online) 
Spin-wave spectrum along high symmetry directions for the modified KH-$J_2$-$J_3$ model under zigzag magnetic order, with parameters in the second column of Table III of Ref. \cite{Hu2015}. The unit of vertical axis (energy) is $\text{meV}\cdot S$, where for ideal $j_{\text{eff}}=1/2$ state $S=1/2$. Inset depicts the Brillouin zone of the Ir honeycomb lattice. High symmetry points are $\Gamma(0,0,0)$, $X(\pi,0,0)$, $M(\pi,\pi,0)$, and $Y(0,\pi,0)$.
}
\label{fig:5}
\end{figure}

From the fitted model parameters one can compute several experimentally relevant properties. Fig. \ref{fig:5} shows the calculated spin-wave spectrum. It has a significant spin gap (about $20.8\text{meV}\cdot S=10.4\text{meV}$) for spin-wave excitations, which can in principle be measured by future inelastic neutron scattering experiments.

In conclusion, we have proposed an alternative moment assignment of the zigzag magnetic order in Na$_2$IrO$_3$ using first-principles calculations. The magnetic moments are along the direction $\bm{g}\approx\bm{a}+\bm{c}$,  forming an angle of 55$^\circ$ with the $\bm{a}$-axis, locating in the cubic $xy$-plane of the IrO$_{6}$ octahedron, and pointing to the middle of the O-O edge. The $\bm{g}$-configuration is explained by  a modified nnKH model, where additional anisotropic interactions are included.  In our picture, first-principles calculations, the modified nnKH model, and experimental measurements become consistent with each other. Therefore, although more experiments are still needed to distinguish between our $\bm{g}$-configuration and former established $\bm{a}$-configuration, our prediction are highly probable to be supported by future experiments. Spin-wave spectrum is calculated, where the calculated spin gap can in principle be measured by future inelastic neutron scattering experiments.


We would like to emphasize that our proposal (that magnetic moment in NaIrO$_3$ lies along the $\bm{g}$-direction) is also consistent with all known experimental evidence. The most relevant experimental signature to the moment direction is the resonant x-ray magnetic scattering measurements~\cite{Liu2011}, in which the original analysis on the experimental data proposed the ordered moment to be along the $\bm{a}$-axis. 
The same experimental data in Ref.~\cite{Liu2011} has been reanalyzed in Ref. \cite{Lovesey2012}, suggesting that the direction of magnetization makes an angle with the $\bm{c}$-axis about $\omega= 118 ^\circ$ in the $ac$-plane.
Since the angle enclosed by the $\bm{c}$-axis and the the $\bm{a}$-axis is $\beta=109^\circ$, which is very close to 118$^\circ$, it was further proposed that magnetic moments were almost parallel to the $\bm{a}$-axis. 
It is however crucial to realize that the procedure used by these authors to fit the scattering intensity does not distinguish between the $\pm\omega$. 
The angle subtended by the $\bm{c}$-axis and the direction of $-\bm{g}$ is 126$^\circ$, which is also very close to 118$^\circ$.
Note that the moment assignment of $-\bm{g}$ is equivalent to $\bm{g}$ since the zigzag configuration is an AFM state. 
Therefore, we conclude that the $\bm{g}$-direction is indeed an alternative explanation of the experimental data.

The authors acknowledge support from National Science Foundation of China (Grant Nos.  11174009 and 11374018) and National Key Basic Research Program of China (Grant Nos. 2011CBA00109 and 2014CB920902).


\section{\large{Supplemental Material}}
\renewcommand{\thefigure}{S\arabic{figure}}
\setcounter{figure}{0}

\subsection{FM stacking versus AFM stacking of Iridate honeycomb planes}
\label{sec: Fm&Afm}

In Fig. 3(a), the magnetic coupling between Ir honeycomb lattices is illustrated as AFM, according to resonant x-ray magnetic scattering measurements \cite{Liu2011_sm}.  However, the KH model neglects the weak coupling between Ir honeycomb lattices, i.e., considering only one layer of Ir atoms. To be consistent with the KH model, the unit cell in our first-principles calculations also contains only one layer of Ir atoms, which means the stacking order of Ir honeycomb lattices are FM. We consider such consistency reasonable since the parameters of the modified nnKH model, which is adopted to explain the $\bm{g}$-configuration of moment assignment of Na$_2$IrO$_3$ in this Letter,  are extracted from the results of first-principles calculations.

For completeness, it is necessary to check the difference between ferromagnetically and antiferromagnetically coupled Ir honeycomb lattices. For the experimental structure, our first-principles calculations shows that the total energy of the antiferromagnetically coupled supercell is lower than that of the ferromagnetically coupled supercell by 2 meV, which is very small since the supercell contains 8 Ir atoms (totally 48 atoms). On one hand, the small energy difference confirms that the coupling between Ir honeycomb lattices is indeed very weak, supporting the consideration of only one layer of Ir atoms in the KH model. On the other hand, it explains the experimentally observed AFM coupling between Ir honeycomb lattices, which is the configuration for the exact ground state. Moreover, our calcucation shows that the total magnetic moment of the ground state for the AFM stacking configuration indeed  points along the direction $\bm{g}\approx\bm{a}+\bm{c}$.  

The relaxed structure for the AFM stacking is also checked by studying the anisotropic energy by first-principles calculations.  The result are almost the same as that for the experimental structure.  Actually we find that the relaxed structure is almost the same as the experimental structure. While the relaxed structure for the FM stacking, the moment direction of the ground state $\bm{g}$ forms an angle of $70^\circ$ with the $\bm{a}$-axis, deviating from the direction $\bm{a}+\bm{c}$ by $15^\circ$. Accordingly, the lattice parameters $a$, $b$, and $c$ change slightly $1.22\%$, $0.99\%$, and $-0.44\%$, respectively. The deviation of the $\bm{g}$-direction indicates that its sensitivity to structure deviations.

\subsection{Robustness of Coulomb repulsion $U$}

\begin{figure}
\includegraphics[width=.5\columnwidth]{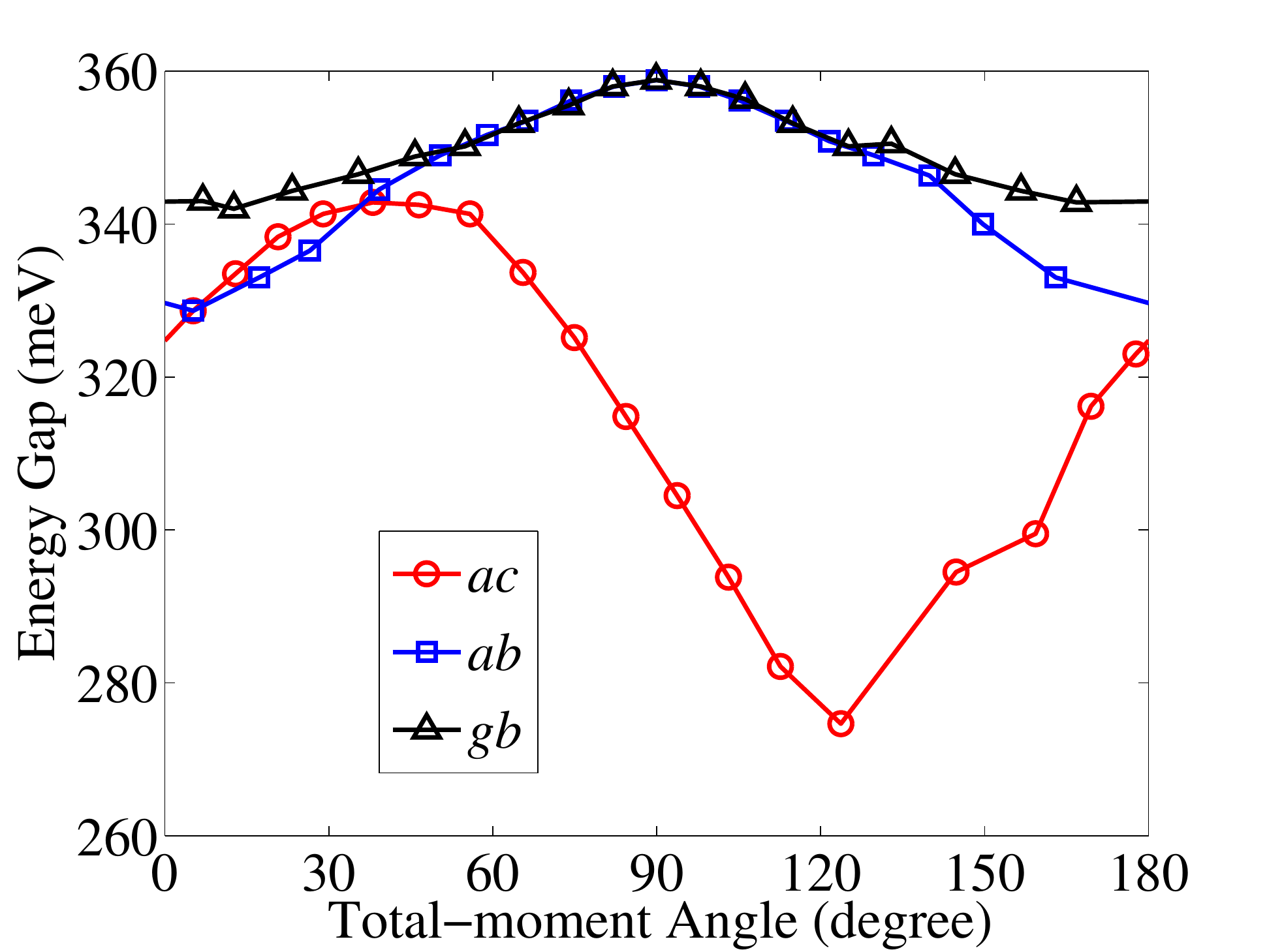}
\caption{(Color online)  Band gap versus total-moment direction. Total-moment angles are measured from the $\bm{a}$-axis to the moment direction for the $ac$- and $ab$-plane, and from the $\bm{g}$-direction for the $gb$-plane, respectively.}
\label{fig:S1}
\end{figure}

The value of Coulomb repulsion $U=1.7$ eV  is chosen carefully to reproduce the band gap provided by experiment \cite{Comin2012_sm}. Fig. \ref{fig:S1} shows the anisotropic band gaps corresponding to different moment directions in $ac$-, $ab$-, and $gb$- planes for $U=1.7$ eV. Band gap varies with moment direction, especially in the $ac$-plane. Various values of $U$ are tested and the $\bm{g}$-configuration of the ground moment assignment turns out to be robust. 

\subsection{Fitting experimental data} 
\label{sec:fitting}

In this section we summarize our fitting results 
for model parameters in Eq.~(2) in main text. Several different fitting schemes are employed.
They differ in the following aspects.

The first factor we consider is
whether to treat spins $S_i^a$ as constant-length-$S$ vectors,
or variable-length vectors with lengths determined by the DFT obtained moment size.
This concern comes from the fact that the moment sizes
do depend on the constrained moment direction, and also the different magnetic orders.

The second factor is 
whether to use all the data available,
or only the low energy states in DFT calculation.
The high energy states in
Fig.~2(b) and Fig. 4(b), 
have worse convergence in DFT calculations 
compared to the low energy ones.
This computational difficulty causes some irregularities in the energy curves in those figures.

The last factor 
is whether to include second-neighbor and third-neighbor interactions.

The fitting results are summarized in
Tables~\ref{tab:1}-\ref{tab:3}.
Note that the fit with nearest-neighbor model (Table~\ref{tab:2})
does not satisfy the condition for 
zigzag ground state.
The second- and third-neighbor Heisenberg couplings
$J_2$ and $J_3$ are thus included in the model,
and their fitting results are presented in Table~\ref{tab:3}.

\begin{table}[h]
\begin{tabular}{|l|c|c|c|c|}
\hline
\multirow{2}{*}{} &
\multicolumn{2}{|c|}{with moment size (meV$/ g^2S^2$)}
&
\multicolumn{2}{|c|}{normalized moments (meV$/ S^2$)} \\
\cline{2-5}
& all data & low energy data & all data & low energy data
\\
\hline
$K$ 
& $-14.4(0.1)$ & $-13.2(0.2)$ & $-5.80(0.08)$ & $-4.96(0.08)$\\
\hline
$J_{\perp\perp}$ 
& $2.0(0.2)$ & $1.0(0.2)$ & $1.0(0.1)$ & $0.45(0.09)$\\
\hline
$J_{\parallel\perp}$ 
& $-1.7(0.2)$ & $-2.2(0.1)$ & $-0.78(0.10)$ & $-0.97(0.05)$\\
\hline
\end{tabular}
\caption{
Fit to the data presented in Fig.~4(b). 
Here ``low energy data'' means 
data below ``0 meV'' in the figure.
The Heisenberg coupling $J$ cannot be reliably fitted 
from these data for zigzag magnetic order only.
Numbers in brackets are estimated error bar 
from the standard least square fit procedure.
Units are $\mathrm{meV}/ g^2S^2$ if
moment sizes are considered, 
where $g$ is the unknown Land\'e $g$-factor;
or $\mathrm{meV}/ S^2$ if moment sizes are normalized.
Ideal $j_{\text{eff}}=1/2$ states will 
have $S=1/2$ and $g=-2$.
}
\label{tab:1}
\end{table}

\begin{table}[h]
\begin{tabular}{|l|c|c|c|c|}
\hline
\multirow{2}{*}{} &
\multicolumn{2}{|c|}{with moment size (meV$/ g^2S^2$)}
&
\multicolumn{2}{|c|}{normalized moments (meV$/ S^2$)} \\
\cline{2-5}
& all data & low energy data & all data & low energy data
\\
\hline
$J$
& $7.2(1.6)$ & $7.2(1.1)$ & $2.8(0.6)$ & $2.7(0.5)$\\
\hline
$K$ 
& $-11.3(2.1)$ & $-9.2(1.5)$ & $-4.9(0.8)$ & $-4.0(0.6)$\\
\hline
$J_{\perp\perp}$ 
& $5.3(2.0)$ & $5.7(1.5)$ & $1.8(0.8)$ & $2.1(0.6)$\\
\hline
$J_{\parallel\perp}$ 
& $-5.2(1.3)$ & $-5.4(0.9)$ & $-2.2(0.5)$ & $-2.1(0.4)$\\
\hline
\end{tabular}
\caption{
Fit to the data presented in Fig.~2(b)
using only nearest-neighbor interactions.
Here ``low energy data'' means 
data below ``30 meV'' in the figure.
}
\label{tab:2}
\end{table}

\begin{table}[h]
\begin{tabular}{|l|c|c|c|c|}
\hline
\multirow{2}{*}{} &
\multicolumn{2}{|c|}{with moment size (meV$/ g^2S^2$)}
&
\multicolumn{2}{|c|}{normalized moments (meV$/ S^2$)} \\
\cline{2-5}
& all data & low energy data & all data & low energy data
\\
\hline
$J$
& $7.2(0.7)$ & $6.7(0.4)$ & $2.8(0.2)$ & $2.7(0.1)$\\
\hline
$K$ 
& $-19.1(0.7)$ & $-16.7(0.6)$ & $-7.1(0.3)$ & $-6.4(0.2)$\\
\hline
$J_{\perp\perp}$ 
& $1.5(0.8)$ & $1.5(0.5)$ & $0.8(0.2)$ & $0.8(0.2)$\\
\hline
$J_{\parallel\perp}$ 
& $-3.5(0.5)$ & $-3.3(0.3)$ & $-1.7(0.2)$ & $-1.5(0.1)$\\
\hline
$J_2$ 
& $-1.6(0.4)$ & $-0.4(0.3)$ & $-0.4(0.1)$ & $0.02(0.10)$\\
\hline
$J_3$ 
& $7.8(0.4)$ & $6.4(0.3)$ & $2.7(0.1)$ & $2.3(0.1)$\\
\hline
\end{tabular}
\caption{
Fit to the data presented in Fig.~2(b)
with second-neighbor and third-neighbor Heisenberg couplings $J_2$ and $J_3$,
in addition to Eq.~(2) in main text.
Here ``low energy data'' means 
data below ``30 meV'' in the figure.
}
\label{tab:3}
\end{table}

From the above results, we see that 
the ferromagnetic Kitaev interaction is always dorminant,
independent with the fitting scheme we use.
We believe that this is the robust conclusion we can reach 
from this analysis.

\subsubsection*{Some analytic results about the modified Kitaev-Heisenberg model Eq.~(2)}
\label{sec:analytic}

The classical ground states of model Eq.~(2) in main text has been 
numerically studied by Rau and Kee in Ref.~\cite{Rau2014arxiv_sm}.
Here we report some analytic results about classical ground state energy
under the four possible magnetic ordering patterns.
\begin{itemize}
\item
Zigzag states:
the classical ground state energy per site is
\begin{equation*}
E_{\text{zigzag}}/S^2
=
\frac{J}{2}-J_2-\frac{3J_3}{2}
-\frac{J_{\perp\perp}}{4}
+\frac{J_{\parallel\perp}}{2}
-\sqrt{
\left (
\frac{-4K+J_{\perp\perp}-2J_{\parallel\perp}}{4}
\right )^2
+\frac{J_{\perp\perp}^2}{2}
},
\end{equation*}
when 
the moments are along 
$\pm (\frac{\sin\theta_Z}{\sqrt{2}},
\frac{\sin\theta_Z}{\sqrt{2}},
\cos\theta_Z)$,
and $\theta_Z$ satisfies 
\begin{eqnarray*}
\cos(2\theta_Z) & = & -\frac{-4K+J_{\perp\perp}-2J_{\parallel\perp}}{
\sqrt{
\left (
-4K+J_{\perp\perp}-2J_{\parallel\perp}
\right )^2
+8J_{\parallel\perp}^2
}}
,\\
\sin(2\theta_Z) & = & -\frac{2\sqrt{2}J_{\parallel\perp}}{
\sqrt{
\left (
-4K+J_{\perp\perp}-2J_{\parallel\perp}
\right )^2
+8J_{\parallel\perp}^2
}}.
\end{eqnarray*}

\item
Stripy state:
the classical ground state energy per site is
\begin{equation*}
E_{\text{stripy}}/S^2
=
-\frac{J}{2}-J_2+\frac{3J_3}{2}
+\frac{J_{\perp\perp}}{4}
-\frac{J_{\parallel\perp}}{2}
-\sqrt{
\left (
\frac{-4K+J_{\perp\perp}-2J_{\parallel\perp}}{4}
\right )^2
+\frac{J_{\perp\perp}^2}{2}
},
\end{equation*}
when 
the moments are along 
$\pm (\frac{\sin\theta_S}{\sqrt{2}},
\frac{\sin\theta_S}{\sqrt{2}},
\cos\theta_S)$,
and $\theta_S$ satisfies 
\begin{eqnarray*}
\cos(2\theta_S) & = & \frac{-4K+J_{\perp\perp}-2J_{\parallel\perp}}{
\sqrt{
\left (
-4K+J_{\perp\perp}-2J_{\parallel\perp}
\right )^2
+8J_{\parallel\perp}^2
}}
,\\
\sin(2\theta_S) & = & \frac{2\sqrt{2}J_{\parallel\perp}}{
\sqrt{
\left (
-4K+J_{\perp\perp}-2J_{\parallel\perp}
\right )^2
+8J_{\parallel\perp}^2
}}.
\end{eqnarray*}

\item
N\'eel state:
the classical ground state energy per site is
\begin{equation*}
E_{\text{N\'eel}}/S^2
=
-\frac{3J}{2}+3J_2-\frac{3J_3}{2}-K
+\left \{
\begin{array}{ll}
-(J_{\perp\perp}+2J_{\parallel\perp}),
&
J_{\perp\perp}+2J_{\parallel\perp}>0;\\
\frac{1}{2}
(J_{\perp\perp}+2J_{\parallel\perp}),
&
J_{\perp\perp}+2J_{\parallel\perp}<0.
\end{array}
\right.
\end{equation*}
The moments will be along 
$\pm \frac{1}{\sqrt{3}}(1,1,1)$ direction 
for the former case($J_{\perp\perp}+2J_{\parallel\perp}>0$),
and 
be along 
$\pm (\sin\theta\cos\phi,\sin\theta\sin\phi,\cos\theta)$
with 
$(\cos2\theta,\sin 2\theta)
=
\frac{(\sin2\phi,-2(\cos\phi+\sin\phi))}{\sqrt{\sin^2 2\phi+4(\cos\phi+\sin\phi)^2}}$
for the latter case($J_{\perp\perp}+2J_{\parallel\perp}<0$).

\item
Ferromagnetic state:
the classical ground state energy per site is
\begin{equation*}
E_{\text{FM}}/S^2
=
\frac{3J}{2}+3J_2+\frac{3J_3}{2}+K
+\left \{
\begin{array}{ll}
(J_{\perp\perp}+2J_{\parallel\perp}),
&
J_{\perp\perp}+2J_{\parallel\perp}<0;\\
-\frac{1}{2}
(J_{\perp\perp}+2J_{\parallel\perp}),
&
J_{\perp\perp}+2J_{\parallel\perp}>0.
\end{array}
\right.
\end{equation*}
The moments will be along 
$\pm \frac{1}{\sqrt{3}}(1,1,1)$ direction 
for the former case($J_{\perp\perp}+2J_{\parallel\perp}<0$),
and 
be along 
$\pm (\sin\theta\cos\phi,\sin\theta\sin\phi,\cos\theta)$
with 
$(\cos2\theta,\sin 2\theta)
=
\frac{(\sin2\phi,-2(\cos\phi+\sin\phi))}{\sqrt{\sin^2 2\phi+4(\cos\phi+\sin\phi)^2}}$
for the latter case($J_{\perp\perp}+2J_{\parallel\perp}>0$).

\end{itemize}

From these results one can see that 
(1) for zigzag state energy to be lower than stripy state energy,
we need  $2J-6J_3-J_{\perp\perp}+2J_{\parallel\perp}< 0$;
(2) for the zigzag state to have moments along local $(1,1,0)$ direction (close to the $\boldsymbol{g}$ direction
in main text),
we need 
$J_{\parallel\perp}\approx 0$,
and 
$-4K+J_{\perp\perp}-2J_{\parallel\perp}>0$.

\subsubsection*{Calculated spin-wave spectrum}
\label{sec:spinwave}

\begin{figure}
\includegraphics[width=0.75\columnwidth]{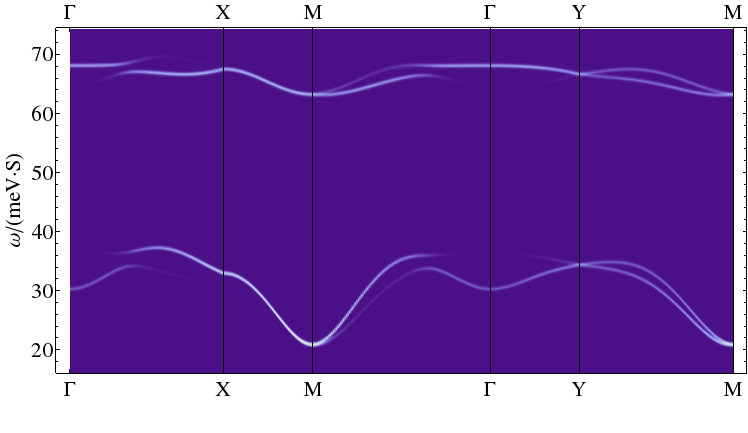}
\caption{(Color online) 
Spin-wave spectrum along high symmetry directions for the modified KH-$J_2$-$J_3$ model,
with parameters in the forth column of Table \ref{tab:3}. The unit of vertical axis (energy) is $\text{meV}\cdot S$, where for ideal $j_{\text{eff}}=1/2$ state $S=1/2$.
Brighter region has larger spectral weight.
}
\label{fig:S2}
\end{figure}

Spin-wave spectrum shown in Fig. 5 in main text and Fig.~\ref{fig:S2} is calculated by the linear spin-wave theory 
using the fitting parameters in the second column of Table~\ref{tab:3}. 
The magnetic moment direction is determined by the solution in last section. 
In fact the spin gap $\Delta_{\text{SW}}(M)$ at $M$ point 
under zigzag magnetic order can be solved analytically, which reads
\begin{equation*}
\Delta_{\text{SW}}(M)
=\sqrt{A^2+B^2-C^2-D^2-2\sqrt{A^2 B^2+C^2 D^2-B^2 D^2}},
\end{equation*}
where
\begin{eqnarray*}
A & = & -J+3J_3+\frac{J_{\perp\perp}}{2}-J_{\parallel\perp}
+\frac{1}{2}\sqrt{
(
-4K+J_{\perp\perp}-2J_{\parallel\perp}
)^2
+8J_{\parallel\perp}^2
},
\\
B & = & -\frac{K}{2}-\frac{3J_{\perp\perp}}{4}
+\frac{ 8K^2 -6KJ_{\perp\perp}+J_{\perp\perp}^2+4K J_{\parallel\perp}-10 J_{\perp\perp}J_{\parallel\perp} }{
4 \sqrt{
(
-4K+J_{\perp\perp}-2J_{\parallel\perp}
)^2
+8J_{\parallel\perp}^2
}
},
\\
C & = & -J+3J_3-\frac{K}{2}+\frac{J_{\perp\perp}}{4}
+\frac{ 8K^2 -6KJ_{\perp\perp}+J_{\perp\perp}^2+4K J_{\parallel\perp}-10 J_{\perp\perp}J_{\parallel\perp} }{
4 \sqrt{
(
-4K+J_{\perp\perp}-2J_{\parallel\perp}
)^2
+8J_{\parallel\perp}^2
}
},
\\
D & = & 
\sqrt{2} K
\sqrt{1-\frac{-4K+J_{\perp\perp}-2J_{\parallel\perp} }{ \sqrt{
(
-4K+J_{\perp\perp}-2J_{\parallel\perp}
)^2
+8J_{\parallel\perp}^2
}
}} 
\\ & &
+(J_{\parallel\perp}-J_{\perp\perp})
\sqrt{1+\frac{-4K+J_{\perp\perp}-2J_{\parallel\perp} }{ \sqrt{
(
-4K+J_{\perp\perp}-2J_{\parallel\perp}
)^2
+8J_{\parallel\perp}^2
}
}}.
\end{eqnarray*}

\end{document}